\documentclass[runningheads]{llncs}
\usepackage{graphicx} 
\usepackage[frozencache=true,cachedir=.]{minted}
\usepackage[hyphens]{url}
\usepackage{caption}
\usepackage{subcaption}

\setminted{breaklines=true}

\begin{document}
\title{Serving Hybrid-Cloud SQL Interactive Queries at Twitter}
%
%

\author{Chunxu Tang \and Beinan Wang \and Huijun Wu \and Zhenzhao Wang \and Yao Li \and Vrushali Channapattan \and Zhenxiao Luo \and Ruchin Kabra \and Mainak Ghosh \and Nikhil Kantibhai Navadiya \and Prachi Mishra \and Prateek Mukhedkar \and Anneliese Lu}
\authorrunning{C. Tang et al.}
%
\institute{Twitter, Inc., San Francisco, United States \\
\email{\{chunxut, beinanw, huijunw, zhenzhaow, yaoli, vrushali, zluo, rkabra, mghosh, nnavadiya, prachim, pmukhedkar, anneliesel\}@twitter.com}}

\maketitle              
\begin{abstract}
The demand for data analytics has been consistently increasing in the past years at Twitter. In order to fulfill the requirements and provide a highly scalable and available query experience, a large-scale in-house SQL system is heavily relied on. Recently, we evolved the SQL system into a hybrid-cloud SQL federation system, compliant with Twitter’s Partly Cloudy strategy. The hybrid-cloud SQL federation system is capable of processing queries across Twitter’s data centers and the public cloud, interacting with around 10PB of data per day.

In this paper, the design of the hybrid-cloud SQL federation system is presented, which consists of query, cluster, and storage federations. We identify challenges in a modern SQL system and demonstrate how our system addresses them with some important design decisions. We also conduct qualitative examinations and summarize instructive lessons learned from the development and operation of such a SQL system.
\end{abstract}

\section{Introduction}\label{sec.intro}

Twitter runs multiple large Hadoop clusters of over 300PB of data, which are among the biggest in the world \cite{twitter-hadoop}. Billions of events are ingested into these clusters per minute \cite{vijayarenu2020scaling}. Twitter's data platform operates these clusters and exerts significant effort in pursuing system scalability and availability to fulfill the data analytics on such large volume data inventory and high throughput data flow. Data customers send tens of thousand queries for data analytics on this huge amount of data daily, usually in SQL statements.

\begin{figure}[htb]
	\centering
	\includegraphics[scale=0.22]{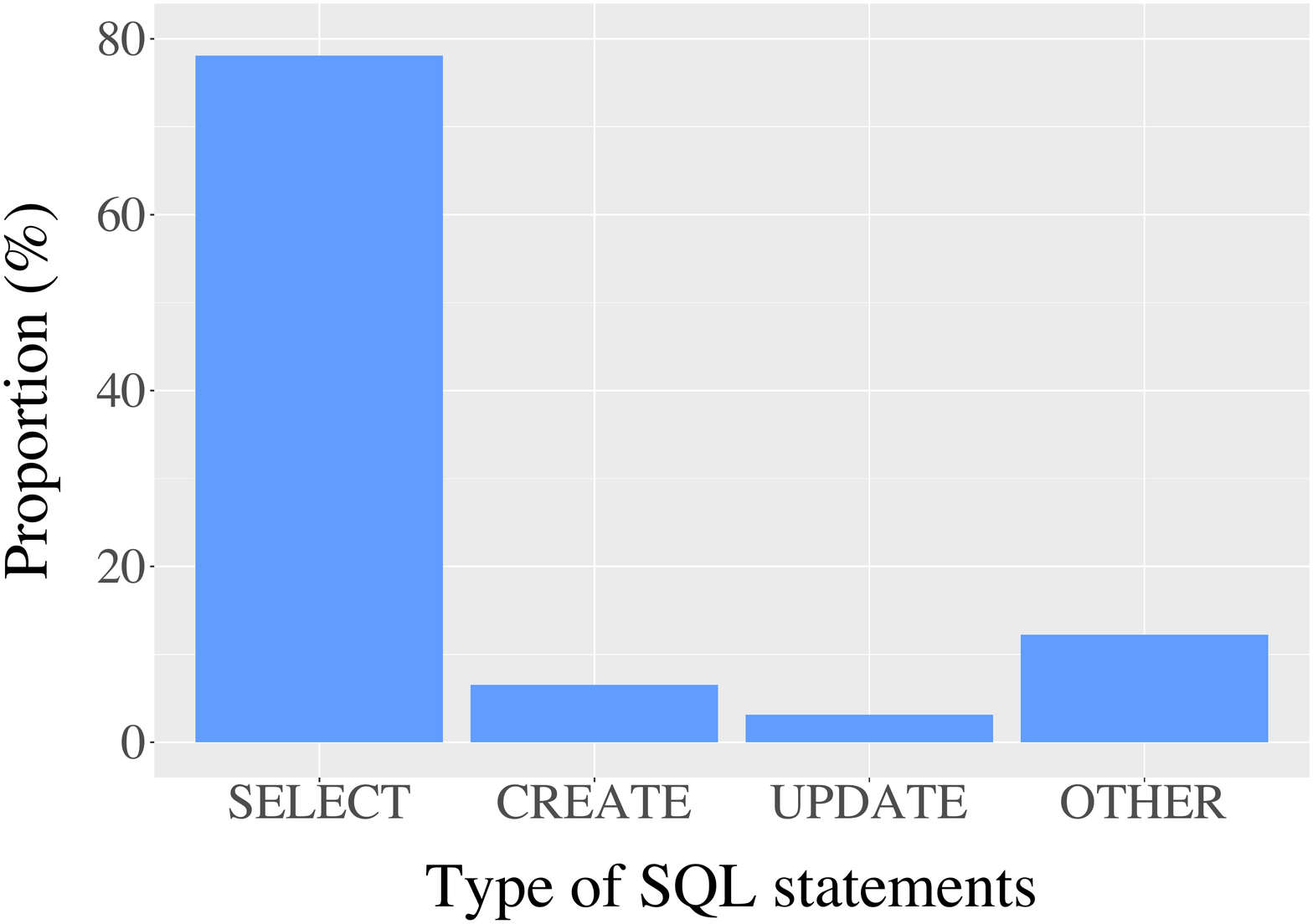}
	\caption{Distribution of SQL statements from a typical Twitter's OLAP workload in a three-month session.}
	\label{fig:sql-type}
\end{figure}

At Twitter, a typical OLAP (Online Analytical Processing) workload mainly contains ad-hoc queries, empowering a wide range of use cases from internal tooling reporting to ads click-rate analysis. The query latencies range from seconds to minutes. There are various query types observed. As Figure \ref{fig:sql-type} indicates, \textit{SELECT} statements dominate the distribution of SQL statements in a typical Twitter's OLAP workload. Users leverage this type of SQL statements to query various datasets stored in the persistent storage. Besides \textit{SELECT} statements, a typical Twitter's OLAP workload also contains \textit{CREATE} statements to create temporary tables or material views, \textit{UPDATE} statements to update records in temporary tables\footnote{At Twitter, SQL system users cannot create or update datasets except exclusive temporary tables under personal accounts. Due to the requirements for data lineage and governance, only data pipeline system accounts have write access to public datasets. SQL individual users only have read access.}, and \textit{OTHER} statements mainly for metadata querying.

Such a SQL system needs to be capable of processing a large number of queries in parallel. Previously, we implemented an in-house SQL system in Twitter's data center (aka private cloud) with hundreds of worker nodes, accompanied by internal Twitter services such as monitoring and logging. At present, to enjoy benefits at a global scale such as faster capacity provisioning, a broader ecosystem of tools and services in the cloud, and enhanced disaster recovery capabilities, Twitter engineering is embarking on an effort to migrate ad-hoc clusters to the GCP (Google Cloud Platform), aka the ``Partly Cloudy'' \cite{partlycloudy}. Partly Cloudy extends Twitter's environment into the public cloud, as a first-class offering alongside on-premises platform services. Under the umbrella of Partly Cloudy, multiple large-scale data analytics jobs or systems \cite{li2021performance,tang2022taming,wu2021migrate,wu2021move} have been migrated to or supported in the cloud. 

The hybrid-cloud environment brings challenges, leading to a fundamental architectural shift for an OLAP system. From our development and operational experience, a modern unified SQL system should handle a series of challenges:

\begin{itemize}
	\item \textbf{Querying heterogeneous data sources in the application layer.} With the growth of the business, more use cases emerged at Twitter, leading to querying heterogeneous data sources, usually processed by different on-premises or cloud query systems with different configurations and interfaces. For example, data scientists from the Health team query data stored in HDFS (Hadoop Distributed File System), processed by HDFS-compatible SQL engines such as Hive \cite{thusoo2009hive}, SparkSQL \cite{armbrust2015spark}, and Presto \cite{sethi2019presto}, to analyze hate speech in the social media platform. Data engineers from the Ads team query data stored in GCS (Google Cloud Storage), processed by cloud query engines such as Presto on GCP, to validate engagement logging data existence and correctness of data schemas. Infrastructure engineers from the Tooling team gain insights from the usage data stored and processed in MySQL and create shareable dashboards. Use cases may also involve querying and joining tables from various data sources. A modern SQL system should support querying heterogeneous data sources in a unified interface.
	
	\item \textbf{Horizontal scaling in the computation layer.} We have witnessed a boost in the number of daily queries processed by Twitter's SQL system in the recent few years. From our operational experience, vertical scaling cannot handle this large number of analytical queries, which can cost a considerable amount of resources\footnote{From an analysis of a typical Twitter OLAP workload in three months, 19.2\% of queries consume more than 1TB peak memory.}. A modern SQL system usually prefers the horizontal scaling approach to serve analytical queries \cite{tan2019choosing}. In addition, as an on-premises data center usually has a limited capacity, the horizontal scaling may need to cross data centers or on-premises/cloud environments. As a result, the SQL system needs to tackle the challenges brought by horizontal scaling, such as cluster orchestration, workload balancing, and fault tolerance. 
	
	\item \textbf{Heterogeneous storage systems in the storage layer.} With the advent of the Big Data era, large-scale storage systems are developed to fulfill the requirements of archiving the scaling volume of data while also maintaining data availability and consistency. The variety of on-premises and cloud data storage systems also poses challenges for a modern SQL system. Operating heterogeneous storage systems is a major challenge we have faced in developing and operating Twitter's SQL system. In a modern SQL system, no matter where the dataset is stored: on-premises storage clusters and/or cloud storage systems, query engines should access the dataset through a unified interface without memorizing the concrete physical paths of target datasets.
\end{itemize}

To overcome these challenges, Twitter engineering teams implement a hybrid-cloud SQL federation system, which processes around 10PB of data daily in production. This paper presents the evolution of the SQL system at Twitter including query federation, cluster federation, and storage federation. It extends our prior work \cite{tang2021hybrid} from multiple perspectives. First of all, we analyze a typical Twitter's OLAP workload, whose query distribution indicates the technical design direction for the SQL federation system. Second, we discuss the rationales behind some design choices in detail such as adopting Zeppelin and Presto for query federation. Third, we explain the unified cluster provisioning strategy with some practical examples, supporting the SQL system deployment across on-premises and cloud environments. Fourth, as a SQL federation system spans a wide range of components, we discuss related work from more perspectives such as data integration and virtualization. Finally, we report some key findings during the development and operation of such a SQL system: Ad-hoc resource-consuming queries are a challenge for scaling a SQL federation system; A centralized hybrid-cloud IAM (Identity and Access Management) can help reduce the technical complexity of implementing IAM across the public and private cloud.

The remainder of this paper is organized as follows. We describe the architectural design and implementation of the hybrid-cloud SQL federation system in Section \ref{sec.arch} to address the aforementioned challenges, discuss related work in Section \ref{sec.related-work}, and reflect on lessons learned in a diverse set of contexts in Section \ref{sec.lessons}. Section \ref{sec.conclusion} concludes the paper.

\section{SQL Federation System Design \& Implementation}\label{sec.arch}

\subsection{Overview}

\begin{figure}[htb]
    \centering
    \includegraphics[scale=0.64]{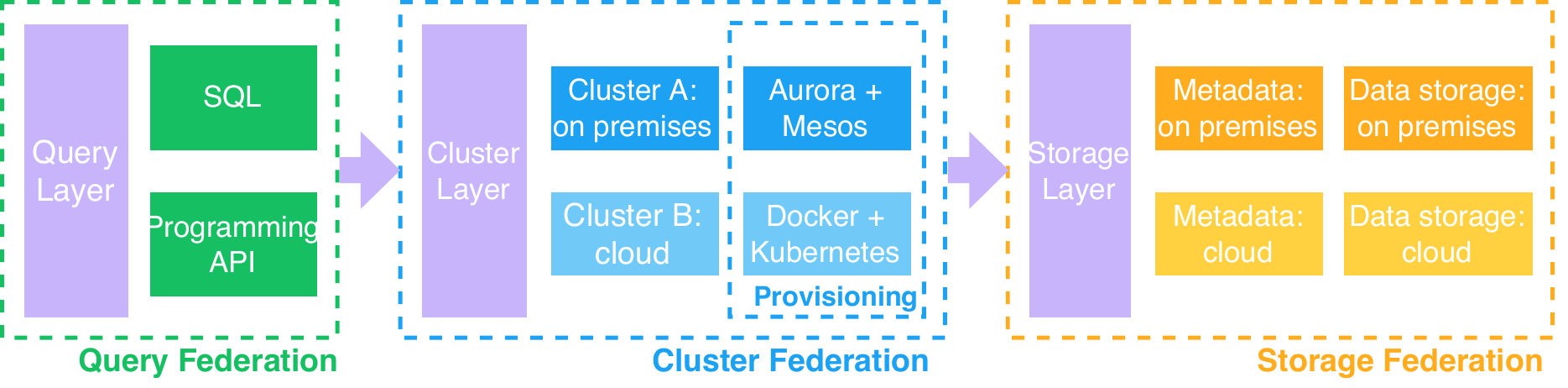}
    \caption{Overview of the hybrid-cloud SQL federation system in three layers.}
    \label{fig:overview}
\end{figure}

Figure \ref{fig:overview} depicts the architectural design of the hybrid-cloud SQL federation system at Twitter. There are three components: query federation, cluster federation, and storage federation. Each federation provides a unified logical view that hides the internal implementation details. This enhances system flexibility and resilience because as long as the logical interface is consistent, any changes in a specific component will cause minimal implementation changes in other federation components.

\textbf{Query federation}. This exposes a unified query layer to customers such that one interface rules multiple query clusters for heterogeneous data sources. Query federation consists of a SQL component and a programming API component. At Twitter, the SQL component supports basic ANSI SQL semantics as well as some Twitter-specific features implemented into UDFs (user-defined functions). The programming API component enables auxiliary flexible programming features. User requests are eventually converted to SQL and passed to the cluster federation.

\textbf{Cluster federation.} This provides a unified cluster layer to the query federation, resolving the challenge of horizontal scaling. It exposes a single entry point, a router service, and hides the cluster details, which reduces the development and operation cost. The router service acts as the coordinator of SQL engine clusters, helping to schedule queries across the clusters and balancing the workloads among the clusters. Fault tolerance is also improved by forwarding requests only to available clusters when a cluster fails and is offline.

\textbf{Storage federation.} This offers a unified view of datasets stored in different archival systems. At Twitter, we are heavily leveraging HDFS as the major on-premises distributed storage platform. In a cloud environment like GCP, we use GCS as the core storage system. The unified layer provides a unique Twitter resolved path for each dataset stored in both on-premises and cloud, entirely getting rid of the burden of memorizing accurate physical locations for datasets.

\subsection{Query Federation}\label{sec.query}

The query federation fulfills three goals. First, as a user-facing front-end, it converts user inputs to SQL and feeds SQL to the cluster federation. Second, it defines datasets in SQL such that users can locate data from different sources with a uniform approach. Third, it provides UI for interaction and visualization. We leverage Zeppelin \cite{zeppelin} to implement the first and third goals, while the second goal is achieved with the help of Presto in the cluster federation. Apache Zeppelin is a web-based notebook service that enables interactive data analytics. Users can fetch data results by sending SQL queries in paragraphs and easily visualize the results with multiple built-in charts or third-party libraries supported by a pluggable framework called Helium \cite{helium}. It should be noted that the SQL front-end design is not limited to Zeppelin, but can be generalized to other notebook tools such as JupyterLab \cite{jupyter}. 

\begin{figure}[htb]
    \centering
    \includegraphics[scale=0.2]{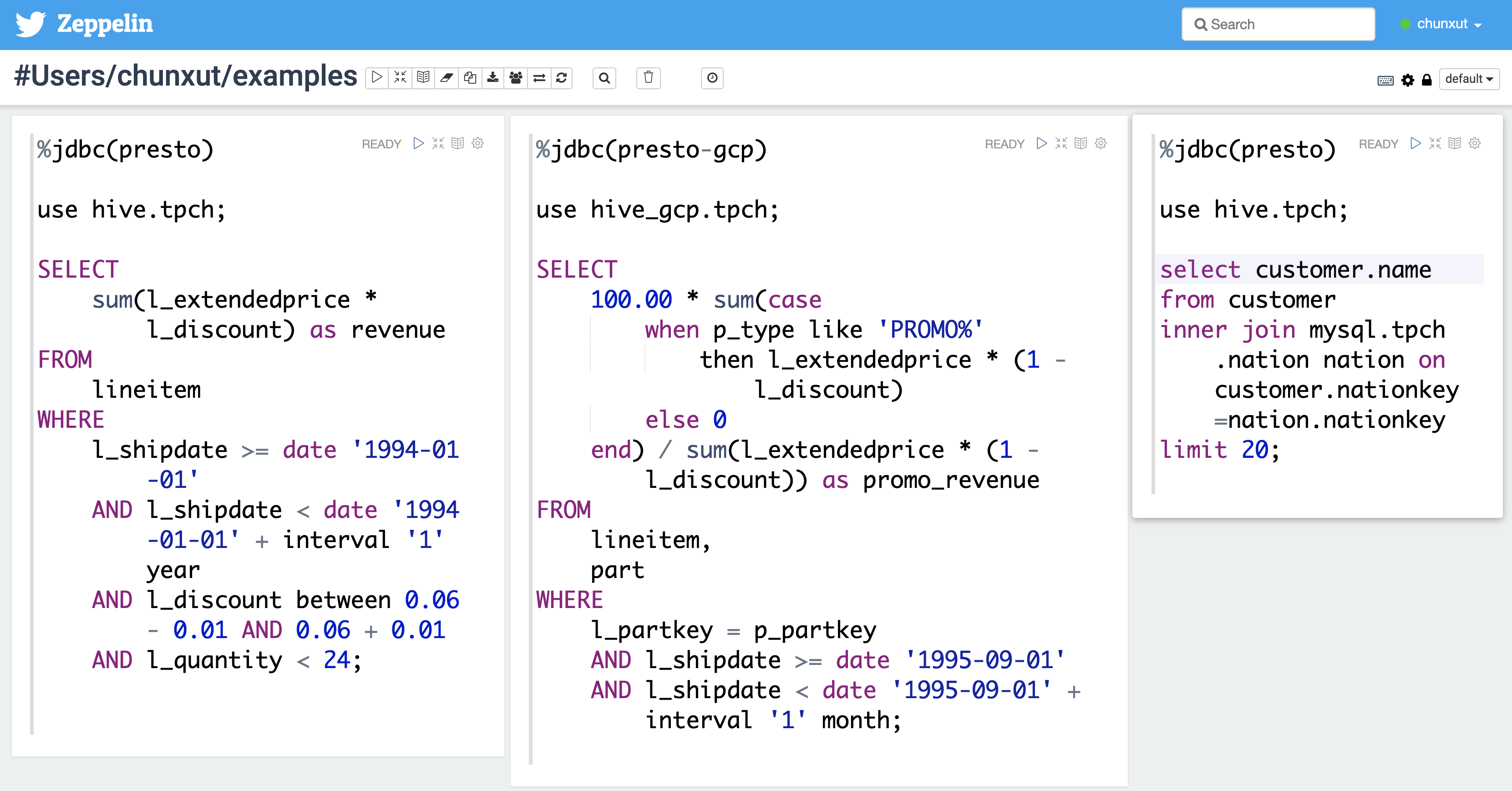}
    \caption{Three SQL query federation examples in a Zeppelin notebook. All SQL statements are from the TPC-H benchmark \cite{tpch}.}
    \label{fig:zeppelin}
\end{figure}

Figure \ref{fig:zeppelin} illustrates some SQL examples of query federation in a Zeppelin notebook. In the figure, the first query and the second one are pointing to the on-premises and cloud SQL clusters respectively, identified by a prefix to flag whether the query should be processed in Twitter's data center or public cloud. No extra configuration is required. Besides explicitly setting dataset locations with prefixes, we are also implementing an automatic data recognition feature by parsing the received SQL statement, extracting target queried tables, and fetching table locality information from Twitter's unified data access layer \cite{dal}. If the dataset is located in the HDFS, the query will be sent to the on-premises SQL engine cluster; if the dataset resides in the cloud, the query will be forwarded to the cloud cluster; if the dataset locates in both HDFS and GCS, we prefer to process the query in the cloud due to more compute resources.

Besides accessing data within one data source, the third SQL statement in Figure \ref{fig:zeppelin} refers to a federated query, joining two tables from HDFS and MySQL. A federated query can refer to joining tables scattered in various data sources. Thus, a query processing engine that can access various data sources should be adopted in the SQL federation system. We adopted Presto for this scenario due to several reasons. First of all, Presto separates compute and storage which allows flexible storage, fitting Twitter's scenario where hundreds of petabytes of data has been stored in HDFS, so it is extremely challenging to migrate data to another storage. Second, Presto is a distributed SQL query engine targeting ``SQL on everything''. With a Connector API communicating with external data stores, data is fetched and then converted to the unified internal Presto data types, so that further query processing, such as joining tables from different data sources, can be accomplished. Lastly, Presto supports interactive queries whose latencies range from seconds to minutes and exposes a web service endpoint, differentiating it from some other query engines such as Hive and Spark which process SQL queries as batch jobs.

\subsection{Cluster Federation}\label{sec.cluster}

\subsubsection{Architectural Design}

Figure \ref{fig:cluster-federation} depicts the architectural design of cluster federation with the following components:

\begin{figure}[htb]
  \centering
  \includegraphics[scale=0.7]{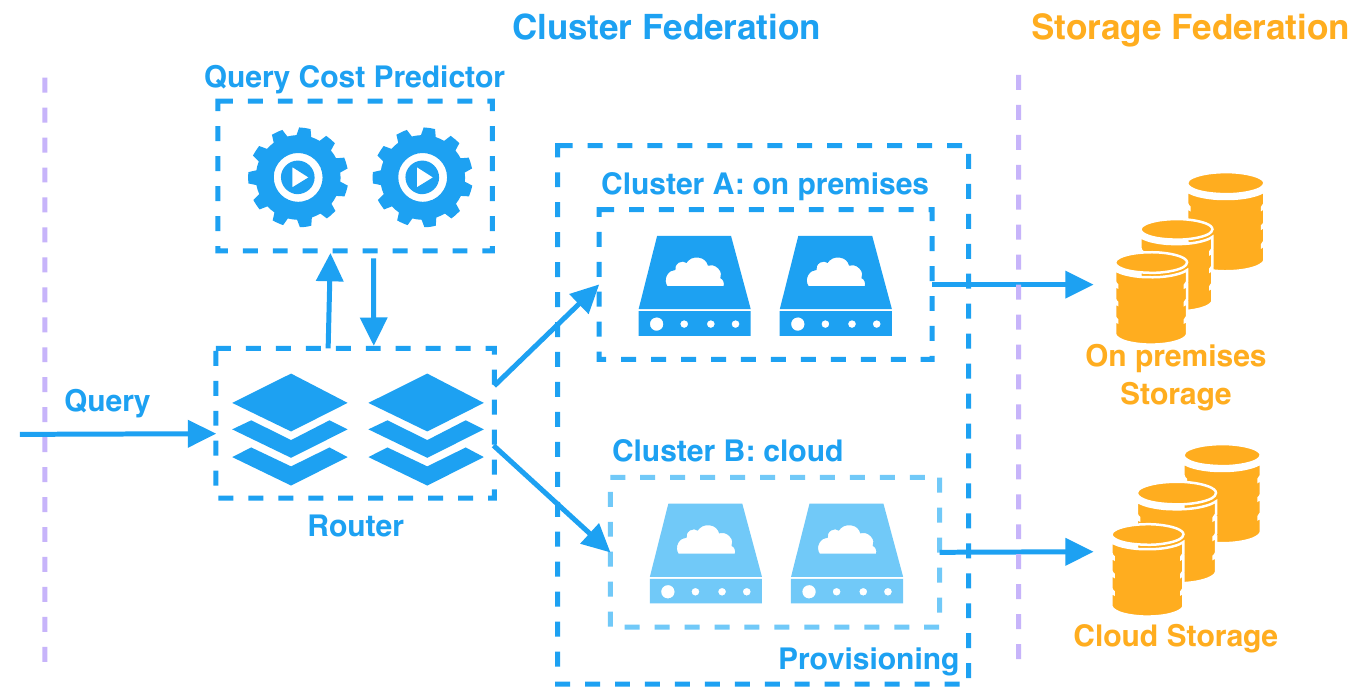}
  \caption{Architectural design of the cluster federation.}
  \label{fig:cluster-federation}
\end{figure}

\textbf{Router.} The router service is the single entry point and the core of the cluster federation, which exposes a unified interface to the query tools, hides cluster details, and routes requests to concrete clusters. Meanwhile, it helps to balance the workloads among the clusters. Our prior SQL system suffered from imbalanced workloads as the clusters were exposed directly to clients. Some clients may send too many queries to a specific cluster, exhausting the compute resources of that cluster, but leaving other clusters idle. The hybrid-cloud SQL federation system harnesses multiple routing algorithms such as round-robin and  random selection. We are also testing more complicated load-based scheduling algorithms with the help of a query cost predictor.

We separate SQL engine cluster endpoint information to a central storage, making the router service stateless. As a result, we can easily scale the service horizontally by deploying multiple router instances. These instances share the same router endpoint and hide proxy details from users. We utilize scheduling algorithms such as the random choice to balance the load on these instances. Even when an instance fails, other router instances can still route requests to SQL engine clusters. This automatic service discovery and recovery improve the availability and avoids the single point of failure on the router.

\textbf{Query cost predictor.} This is a preditor service to forecast the CPU and memory resource usages of each SQL query. It applies machine learning techniques to train models learning from historical SQL queries. The predictor details are beyond the scope of this paper and discussed in a separate paper \cite{tang2021forecasting}.

\textbf{SQL engine cluster.} Presto is the query engine utilized in a SQL engine cluster. Each Presto cluster consists of a coordinator node and one or more worker nodes. A SQL engine cluster may be deployed in Twitter's data center or cloud. When it is deployed in Twitter's data center, it queries data stored in on-premises services such as HDFS. By contrast, when it is in the GCP, it queries data stored in the GCS. The SQL engine clusters do not query data across data centers due to performance concerns.

With the cluster federation, users only view logical clusters. When a cluster fails and is offline, the router will remove it from the available cluster list and will not route any requests to this cluster. When the cluster recovers from the failure and is back online, the router will find the cluster through service discovery, mark it as available, and route requests to this cluster. This also improves the availability and fault tolerance, mitigating the operation pain we have faced in the prior SQL system with separate clusters.

\subsubsection{Cluster Provisioning}

In a hybrid-cloud environment, a service can be deployed in both the private cloud and public cloud. This poses challenges in the product release and deployment as we have to operate two suites of toolchains.

In Twitter's data centers, each service is hosted in an Aurora container, enabling running services and cron jobs on top of Mesos \cite{hindman2011mesos}, a distributed system kernel. Mesos abstracts compute resources away from specific machines to treat the data center like one big computer. Each job is defined and described in Aurora DSL \cite{aurora-dsl}. As developers need to upload applications before deployment, Twitter engineers also develop \textit{Packer} \cite{packer}, a package versioning and storage system, to manage the applications.

By contrast, in a cloud platform such as GCP, we wrap each service into a Docker image and deploy it into the Kubernetes platform, aka GKE (Google Kubernetes Engine) in the GCP. Kubernetes offers a declarative job specification, which can be described in YAML. For example, the simplified code snippet shown below refers to creating a SQL system demo in the Kubernetes.


\begin{minted}
[
frame=lines,
framesep=2mm,
baselinestretch=1,
fontsize=\scriptsize,
fontfamily=Times New Roman,
linenos
]
{yaml}

apiVersion: compute.twitter.com/v1
kind: TwitterSetDeployment
metadata:
  namespace: sqlsystem
  name: sqlsystem-devel
spec:
  replicas: 1
  template:
    metadata:
    spec:
      priorityClassName: preemptible
      containers:
      - name: sqlsystem
        image: sqlsystem
        imagePullPolicy: Always
        command:
        - /bin/bash
        - -c
        args:
        - /usr/lib/jvm/java-1.8.0-twitter/bin/java -jar sqlsystem.jar -instance={{mesos.instance}} -admin.port=':8080'
        resources:
          limits:
            cpu: 1000m
            memory: 2Gi
            ephemeral-storage: 2Gi
        ports:
          - containerPort: 8080
            name: http
          - containerPort: 8080
            name: service

\end{minted}

To fill the gaps between the two cluster provisioning strategies and resolve the challenges of provisioning in a hybrid-cloud environment, Twitter engineers implement an abstraction layer on top of Mesos and Kubernetes, only exposing the Aurora DSL to describe jobs deployed in both private and public cloud.

The following is a simplified example of describing the above GCP job in Aurora DSL, where each key-value pair in YAML is translated into Aurora DSL:

\begin{minted}
[
frame=lines,
framesep=2mm,
baselinestretch=1,
fontsize=\scriptsize,
fontfamily=Times New Roman,
linenos
]
{python}
resources = Resources(cpu = 1.0, ram = 2 * GB, disk = 2 * GB)
port='8080'
jobs = [
  BasicTwitterSet(
    cluster = 'cluster1',
    role = 'sqlsystem',
    environment = 'devel',
    name = 'sqlsystem',
    requires = [Announcer()],
    replicas = 1,
    spec = PodSpec(
      priorityClassName='preemptible',
      containers = [
        KubernetesContainer(
          name='sqlsystem',
          image='sqlsystem',
          command=JVMProcess(
            name = 'run_sqlsystem',
            jvm = Java8,
            arguments = '-jar sqlsystem.jar -instance={{mesos.instance}} -admin.port=":%s"' % port,
            resources=resources
          ),
          resources = resources,
          ports = [
            ContainerPort(name = ['http', 'service'], containerPort = port),]
        )]
    ),
  )
]
\end{minted}

This unified job configuration abstraction greatly reduces the operation cost. We only devise one suite of toolchains but with different configuration details for Twitter's data center and public cloud. 

\begin{figure}[htbp]
  \centering
  \includegraphics[scale=0.55]{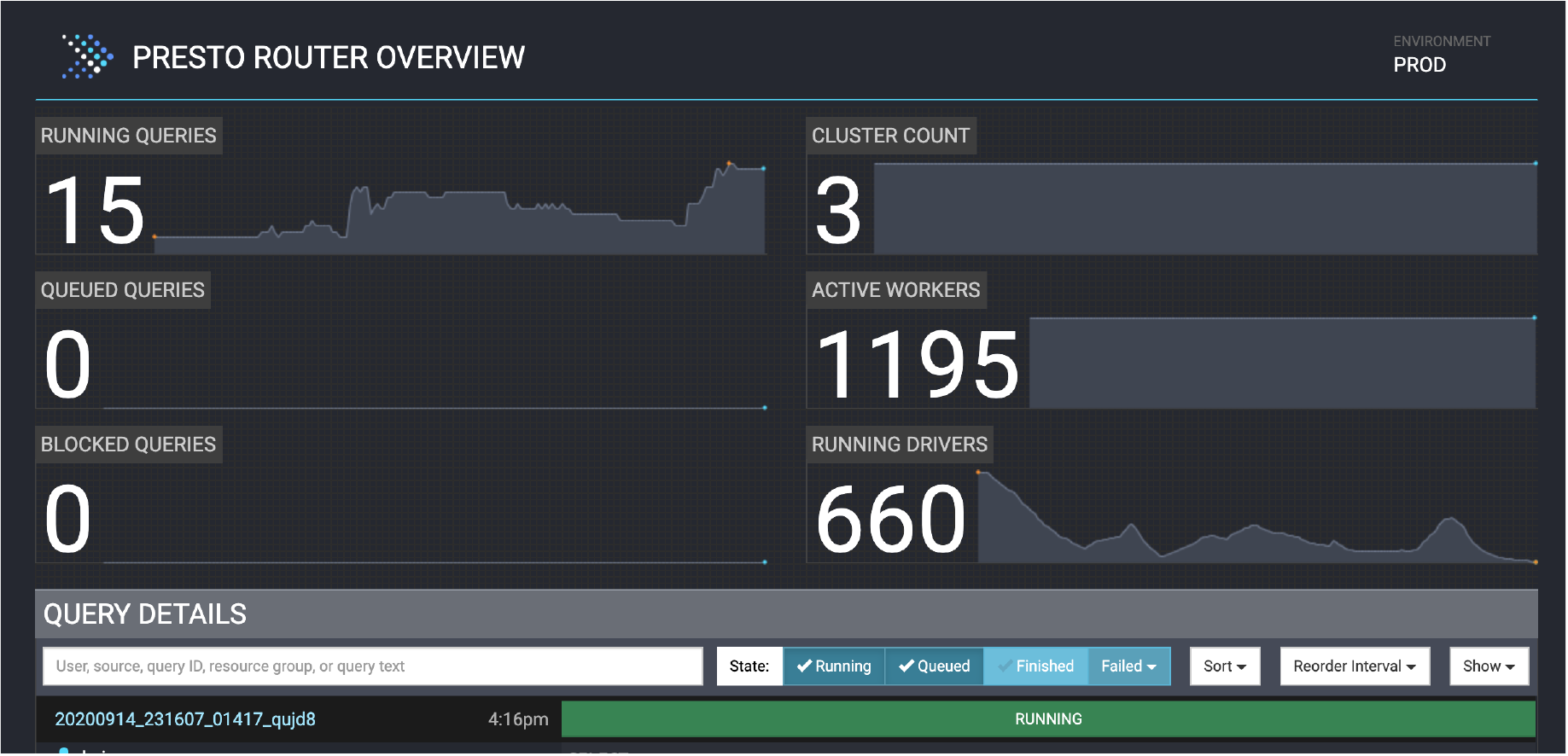}
  \caption{Unified UI for cluster federation.}
  \label{fig:router-ui}
\end{figure}

\subsubsection{Unified Interface}

To ease the administration of SQL engine clusters, we build an aggregated UI, shown in Figure \ref{fig:router-ui}, on top of the original Presto UI. The UI aggregates the status of all SQL engine clusters, sums the running queries, and monitors the active workers. Moreover, we can dive deeper into one specific cluster to investigate the performance metrics, collected into a unified dashboard shown in Figure \ref{fig:presto-mon}. This panel visualizes metrics, including query failures, cluster memory, running queries, etc., collected in the past two weeks.


\newcommand{\monwidth}{0.7\textwidth}
\newcommand{\monheight}{1.5in}

\begin{figure}[htbp]
	\centering
	\begin{subfigure}[t]{\monwidth}
		\centering
		\includegraphics[height=\monheight]{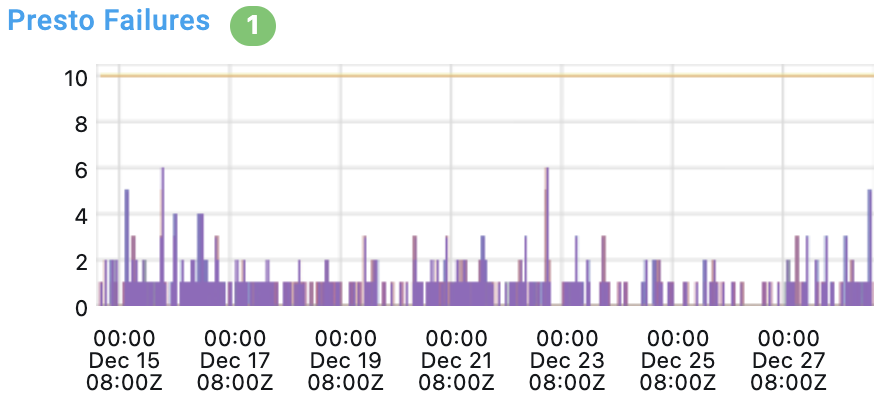}
		\caption{SQL query failures in two weeks.}
		\label{fig:mon-failures}
	\end{subfigure}
	~
	\begin{subfigure}[t]{\monwidth}
		\centering
		\includegraphics[height=\monheight]{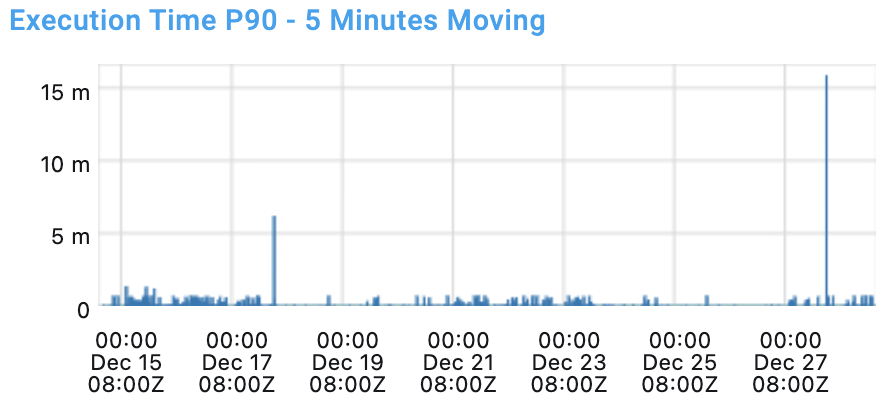}
		\caption{The 5 minutes moving average of execution time P90 (90th percentile) in two weeks.}
		\label{fig:mon-p50}
	\end{subfigure}
  ~
  \begin{subfigure}[t]{\monwidth}
		\centering
		\includegraphics[height=\monheight]{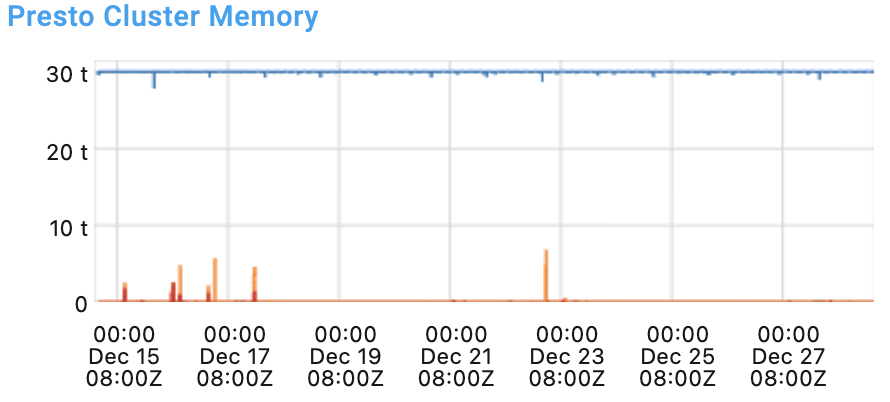}
		\caption{Cluster memory usages in two weeks.}
		\label{fig:mon-cluster-memory}
	\end{subfigure}
  ~
  \begin{subfigure}[t]{\monwidth}
		\centering
		\includegraphics[height=\monheight]{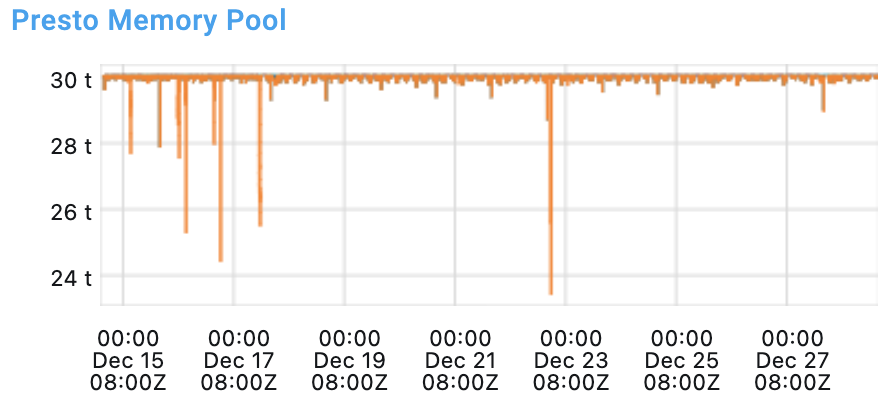}
		\caption{SQL engine memory pool in two weeks.}
		\label{fig:mon-memory-pool}
	\end{subfigure}

	\caption{Monitoring and alerting of one SQL engine (Presto) cluster.}
	\label{fig:presto-mon}
\end{figure}

\subsection{Storage Federation}\label{sec.storage}

\begin{figure}[htb]
    \centering
    \includegraphics[scale=0.68]{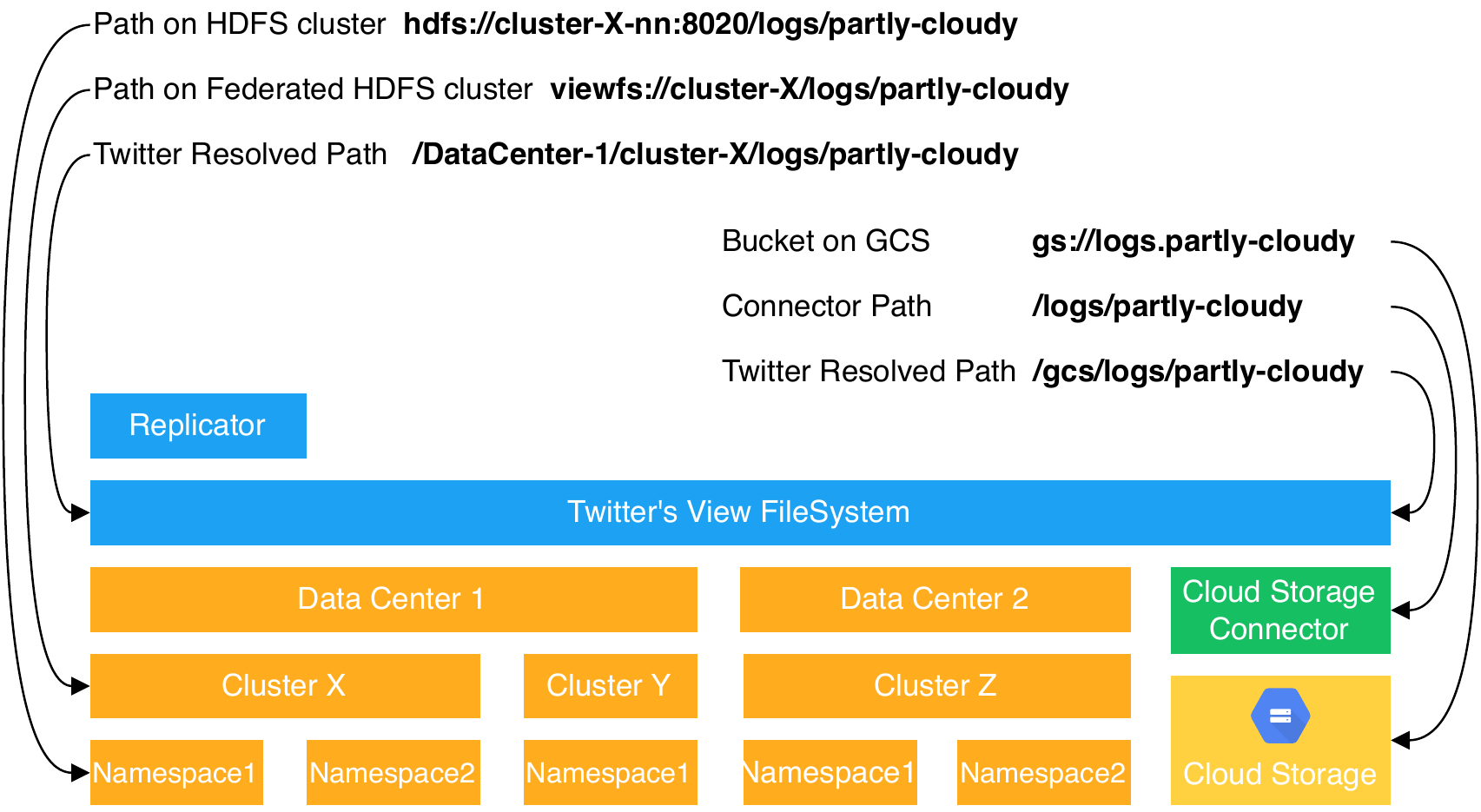}
    \caption{Architectural design of the storage federation.}
    \label{fig:storage-federation}
\end{figure}

To fulfill both scaling data and high availability requirements, Twitter engineers maintain storage clusters in both Twitter's data center and public cloud. Figure \ref{fig:storage-federation} depicts the high-level design of the storage federation platform, which is backed by hundreds of thousands of data replication jobs. This platform contains an unified view for data stored in on-premises HDFS clusters and a cloud storage system (GCS in the GCP).

\begin{figure}[htbp]
	\centering
	\begin{subfigure}[]{0.97\textwidth}
		\centering
		\includegraphics[height=3.7in]{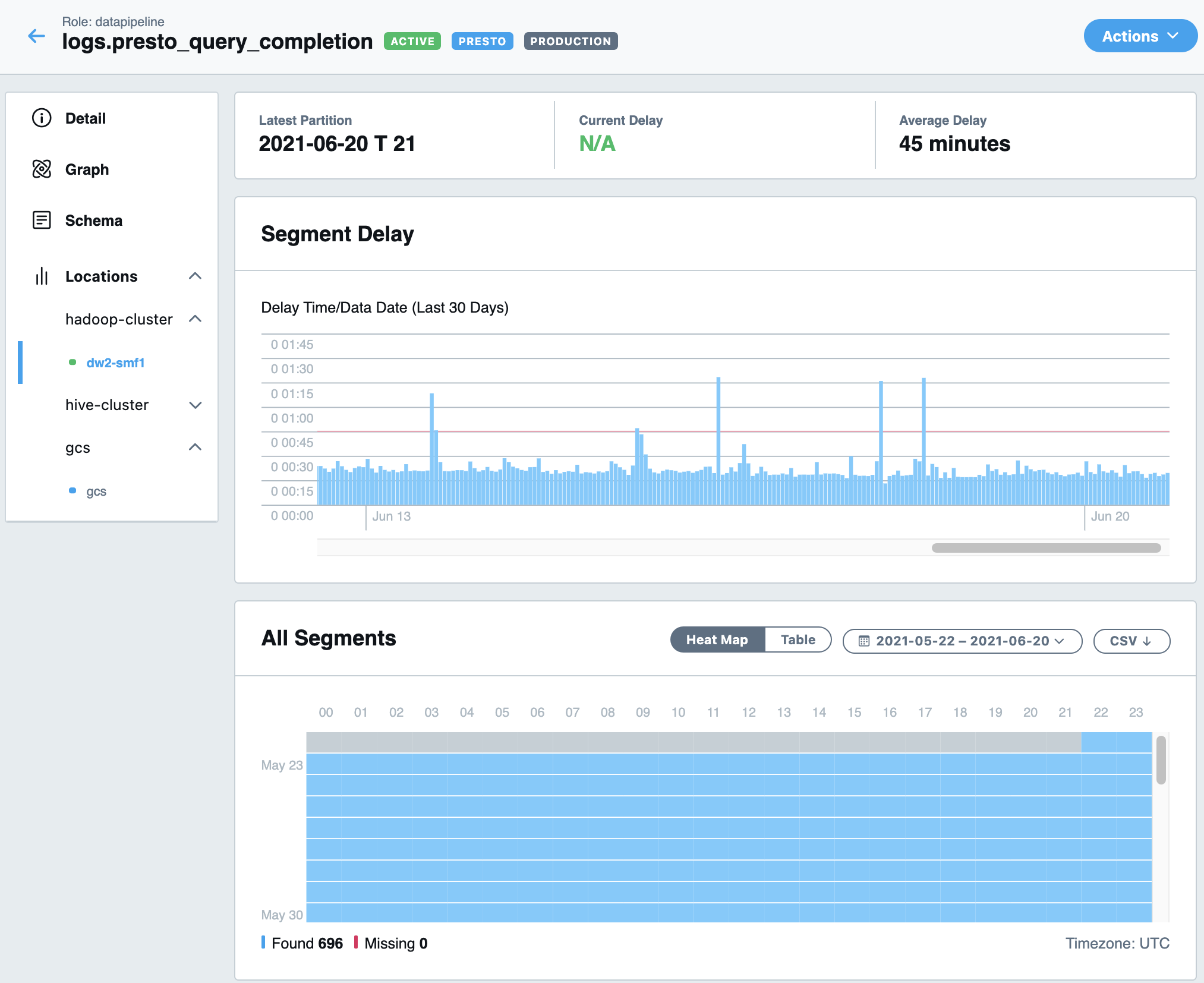}
		\caption{Details of a dataset in on-premises HDFS.}
		\label{fig:logs-hdfs}
	\end{subfigure}
	~
	\begin{subfigure}[]{0.97\textwidth}
		\centering
ain		\includegraphics[height=3.7in]{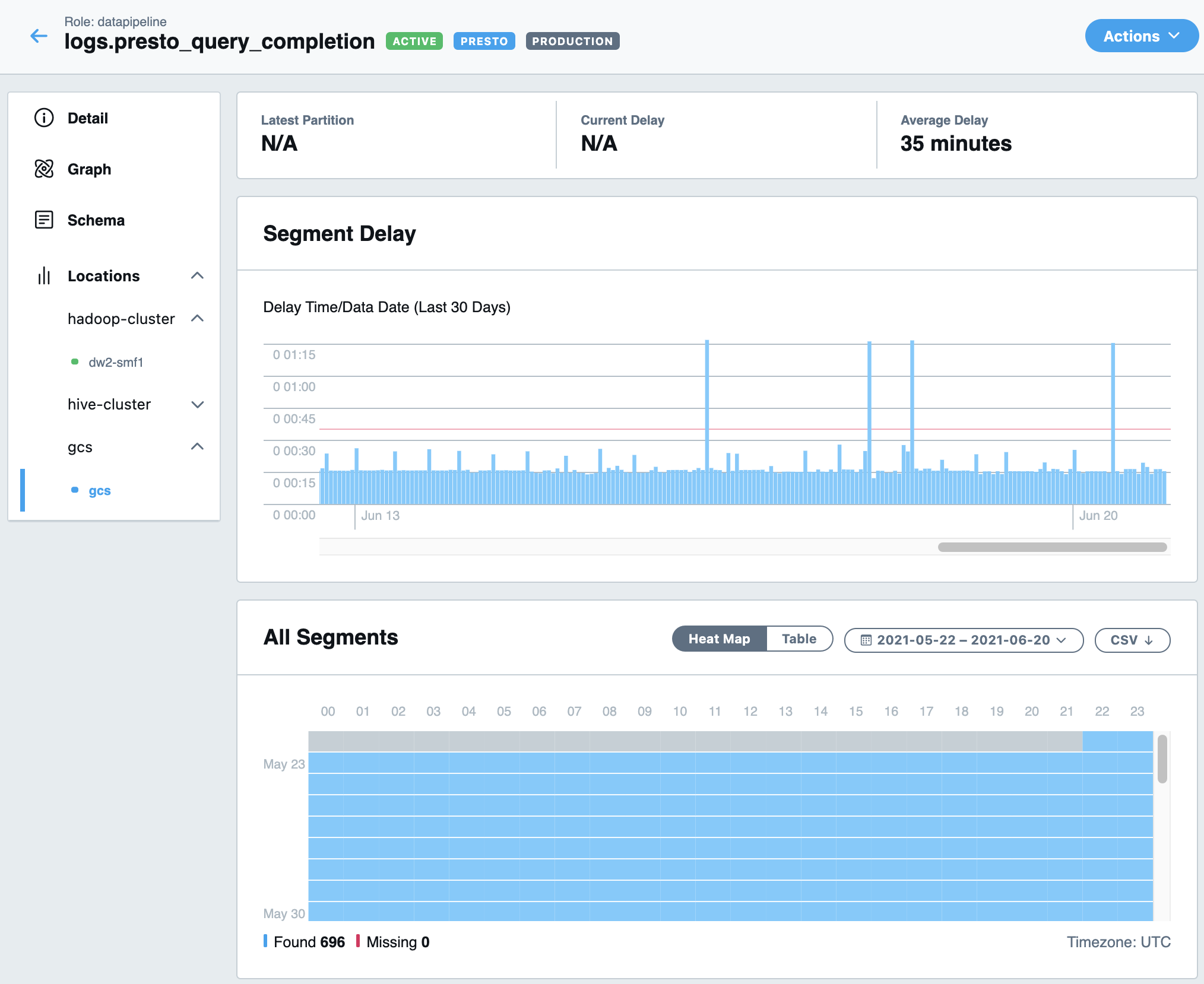}
		\caption{Details of a dataset in GCS.}
		\label{fig:logs-gcs}
	\end{subfigure}
	
	\caption{Unified UI for datasets stored in HDFS and GCS.}
	\label{fig:unified-datasets}
\end{figure}

\textbf{On-premises HDFS.} Twitter's data platform operates multiple HDFS clusters across data centers, shown as the left part in Figure \ref{fig:storage-federation}. Multiple namespaces are also required due to scalability and use case isolation requirements. We scale HDFS by federating these namespaces with user-friendly paths instead of long complicated URIs \cite{hadoop-twitter}. As shown in Figure \ref{fig:storage-federation}, first, the original on-premises data path is \textit{hdfs://cluster-X-nn:8020/logs/partly-cloudy} (\textit{nn} refers to the namenode in HDFS), indicating the data resides in Cluster X in Data Center 1, under the namespace \textit{logs}. Second, we leverage Hadoop ViewFs \cite{viewfs} to provide a single view across namespaces, starting with \textit{viewfs://}. So the original path will become \textit{viewfs://cluster-X/logs/partly-cloudy}. Finally, we extend the ViewFs and implement Twitter's View FileSystem, offering a unified user-friendly path (\textit{/DataCenter-1/cluster-X/logs/partly-cloudy} in Figure \ref{fig:storage-federation}) and enabling native HDFS access. A replicator service is also created to help access data stored in different locations.

\textbf{Cloud storage (GCS).} Because of the large data volume and use case isolation, we maintain thousands of GCS buckets at Twitter. We also leverage the View FileSystem abstraction to hide GCS details behind the storage interface. The cloud storage connector is utilized to interact with GCS via Hadoop APIs. We apply the RegEx-based path resolution to resolve the GCS bucket path, by dynamically creating mountable mapping on-demand in Twitter's View FileSystem. As shown in Figure \ref{fig:storage-federation}, similar to HDFS, the GCS bucket \textit{gs://logs.partly-cloudy} is finally resolved as \textit{/gcs/logs/partly-cloudy}.

As a result, the storage federation only exposes standard unique paths of datasets, no matter they reside in the on-premises HDFS clusters or GCS. In addition, Twitter engineers maintain a metadata service, connected with these storage systems, aiming to identify the closest location of the target dataset and return its standard path to query engines. For example, in Figure \ref{fig:storage-federation}, querying the same \textit{partly-cloudy} dataset, if the query engine is in a Twitter's data center, the on-premises path \textit{/DataCenter-1/cluster-X/logs/partly-cloudy} will be returned. By contrast, if the query engine is in the cloud, the cloud path \textit{/gcs/logs/partly-cloudy} will be returned.

To view dataset configuration details, Twitter engineers create a unified UI, shown in Figure \ref{fig:unified-datasets}, with segment support for files stored in various physical locations. Users can thus view different locations for the same dataset. Specifically, Figure \ref{fig:logs-hdfs} illustrates details of the query log dataset of Presto stored in an on-premises HDFS cluster; Figure \ref{fig:logs-gcs} points to details of the same dataset stored in the GCS. Figures also show segment delays and segment block information.

\section{Related Work}\label{sec.related-work}

A large-scale SQL system involves a wide range of related domains. Here, we discuss related work in each domain.

\subsection{Data Integration and Virtualization}

Data integration involves aggregating data from heterogeneous data stores and in different formats to realize analytics on a huge amount of data. This is usually tackled by either transforming the data to a consistent data format and physically placing data in a data warehouse, or creating a virtualized middleware and offering a unified logical view of datasets, namely data virtualization.

Recently, thanks to advantages such as high scaling capability and on-the-fly processing \cite{mousa2015data}, some research work emerged in the data virtualization domain, supporting ad-hoc queries on heterogeneous data stores. For example, Lawrence \cite{lawrence2014integration} proposed a generic standards-based architecture on top of both SQL and NoSQL systems, verified by MySQL and MongoDB. The virtualization system translates SQL queries into source-specific APIs, with minimal performance overhead reported. The author \cite{lawrence2017faster} later extended this work to support distributed semi-joins. Similarly, Vathy-Fogarassy and Hugy{\'a}k \cite{vathy2017uniform} implemented a uniform data access platform covering heterogeneous data stores including SQL and NoSQL databases. Their solution does not support joins across data sources but only collects data from these data sources. More recently, Mami et al. \cite{mami2019uniform} established a semantic data lake to access and process heterogeneous data at scale. Aleyasen et al. \cite{aleyasen2018high} proposed a context-aware query router, aiming to address the data replication obstacle in on-premises to cloud migration, similar to the router service we implemented in Twitter's data platform. 

Compared with prior work, Twitter's hybrid-cloud SQL federation system relies on both the data warehouse (transforming analytical data to the Parquet \cite{parquet} format and storing them in some centralized storage clusters such as on-premises HDFS and cloud-based GCS) and data virtualization solutions (creating unified logical views for SQL applications, computation, and storage).

\subsection{SQL Systems} 

With the increasing volume of data, many distributed SQL engines, targeted for analyzing Big Data, emerged in the recent decade. For example, Apache Hive \cite{thusoo2009hive} is a data warehouse built on top of Hadoop, providing a SQL-like interface for data querying and a warehousing solution to address some issues of MapReduce \cite{dean2008mapreduce,dean2010mapreduce}. Spark SQL \cite{armbrust2015spark} is a module integrated with Apache Spark, powering relational processing to Spark data structures, with a highly extensible optimizer, Catalyst. Presto \cite{sethi2019presto}, originally developed by Facebook, is a distributed SQL engine, targeting ``SQL on everything''. It can query data from multiple sources which is a major advantage over other SQL engines. It now also supports real-time analytics at scale \cite{luo2022presto}. Twitter's SQL federation system described in this paper chooses Presto as the core SQL engine thanks to its low latency, high flexibility, and high extensibility. Procella \cite{chattopadhyay2019procella} is a SQL query engine, facilitated by YouTube, serving hundreds of billions of queries per day.

With the advent of the public cloud, some cloud-based commercial SQL products rose in popularity in the recent decade. For example, Google BigQuery \cite{bigquery} (a public implementation of Dremel \cite{melnik2010dremel,melnik2020dremel}) offers a cloud-based, fully-managed, and serverless data warehouse. Similarly, Snowflake \cite{dageville2016snowflake} provides a multi-tenant, transactional, and elastic system with full SQL support for both semi-structured and schema-less data. Amazon Redshift \cite{gupta2015amazon} applies a classic shared-nothing architecture with Vertica \cite{lamb2012vertica}-similar compression techniques, acting as a fully-managed PB-scale data warehouse solution in AWS. Azure Synapse Analytics \cite{aguilar2020polaris} separates compute and storage for cloud-native execution, bringing together data warehousing and big data workloads. Alibaba AnalyticDB \cite{zhan2019analyticdb,wei2020analyticdb} offers real-time query processing in hundreds of milliseconds and decouples write and access paths to fit large-scale data analytics.

\subsection{Cluster Management} 

The need for efficient cluster management has led to the creation of various systems in the past decade. YARN \cite{vavilapalli2013apache} is the resource management system used in Hadoop, abstracting cluster management from computation jobs. Google developed a large-scale unified cluster management system, Borg \cite{verma2015large,tirmazi2020borg}, running hundreds of thousands of jobs. It is later open-sourced as Kubernetes. Google also created Omega \cite{schwarzkopf2013omega}, an offspring of Borg to improve the software engineering. Mesos \cite{hindman2011mesos} is a similar resource-sharing platform, used with Aurora configuration Language \cite{aurora-dsl} at Twitter. Facebook established Twine \cite{tang2020twine}, formerly known as Tupperware, to act as a large-scale resource-sharing infrastructure. As Twitter engineering is migrating some service to the GCP, we use both Mesos and Kubernetes for cluster deployment and management.

\section{Lessons Learned}\label{sec.lessons}

In this section, we recount some of the qualitative lessons we have learned from the development and operation of the SQL federation system at Twitter.

\textbf{System monitoring and logging in a hybrid-cloud environment are vital.} Although our hybrid-cloud SQL federation system almost always works well, sometimes when the system goes wrong, it can be a headache to locate the root cause. We also observed architectural differences between on-premises and cloud environments, such as cluster provisioning and security enforcement. An important design decision we have made is implementing a real-time monitoring system with metrics collection and an injectable logging system to trace execution flows. The monitoring system provides a central platform to collect predefined and user-customized metrics, serves observability dashboards/alerts, and helps developers drill down to detailed metrics. Meanwhile, the injectable logging system provides APIs to inject logging points into application source code, collects the logs, and visualizes the execution flows.

\textbf{The on-premises capacity planning experience cannot be directly transferred to a hybrid-cloud environment.} During the migration of parts of on-premises workload to the cloud, we discovered that the capacity planning experience cannot be easily reused and shared across data centers, due to varied technical stacks and resource provisioning strategies. For example, one of our early migrated use cases requires around 50 machines in Twitter's data center but needs around 60 to get comparable performance in the cloud, even though all these machines are sharing similar hardware configuration. Based on the lesson, we suggest additional prototypes for capacity planning and extra tuning of service in a hybrid-cloud environment.

\textbf{Ad-hoc resource-consuming queries are a challenge for scaling a SQL federation system.} In Twitter's data platform, we observed that more than 70\% of interactive SQL queries can be completed in less than 1 minute. But there is a small proportion (around 10\%) of queries that cost lots of compute resources and can cost as long as a few hours to complete. This fact poses challenges in scaling our SQL system, as resource-consuming queries can rapidly exhaust a SQL engine cluster's compute resources. Furthermore, clusters across data centers with different configurations complicate the optimization. We have tried establishing a specific cluster for these resource-consuming queries, implemented multiple scheduling algorithms in the router, and are testing more complicated algorithms with forecasted query resource usages. This also implies a future research direction about more intelligent scheduling on SQL queries with high-variant resource usages across clusters under a hybrid-cloud environment.

\textbf{A centralized hybrid-cloud IAM (Identity and Access Management) can help reduce the technical complexity of implementing IAM across the public and private cloud.} Identity and access management is a crucial component of an enterprise-grade platform including a SQL system. IAM aims to grant the right individuals access to appropriate services. At Twitter, the on-premises SQL system utilizes LDAP (Lightweight Directory Access Protocol) and Kerberos for authorization and authentication. By contrast, GCP dramatically leverages access tokens from OAuth 2.0 for managing identity and access. During the migration, the first solution we raised is to modify codebases of all service components to support GCP IAM. However, later on, we figured out this approach would cost much more complexity in the implementation than expected as we need to explicitly support both LDAP and GSuite logins and management. Given the extreme emphasis on data privacy and security at Twitter, we finally came up with a hybrid-cloud IAM solution with a centralized IAM platform, mapping all Twitter employee LDAP accounts to GSuite accounts with corresponding credentials implicitly, hidden from end-users. This design saved us quarters of engineering effort to set up the IAM in the hybrid-cloud SQL system.

\textbf{SQL is still one of the most widely used languages in data analytics.} As a declarative language, SQL lets users focus on defining the data analytics tasks without worrying about the specifics on how to complete these tasks. Thanks to SQL's high expressiveness in queries and large existing customer bases, some execution engines previously without SQL support, such as Druid \cite{druid-sql} and Beam \cite{beam-sql}, began to support SQL on top of their native query layers. In addition, some SQL variants, such as BigQuery ML \cite{mucchetti2020bigquery}, even introduced SQL into machine learning use cases. From our observation, SQL is still widely used in data analytics, although challenged by some competitive alternatives such as Python. Python is more like a powerful supplement for SQL in data analytics with its concise styles and extreme popularity in machine learning, instead of a complete replacement.

\section{Conclusion}\label{sec.conclusion}

Understanding and identifying challenges faced within a modern SQL system are of ever-growing importance as there is a rapidly growing need for large-scale data analytics. From our development and operational experience at Twitter's data platform, we presented querying heterogeneous data sources in the application layer, horizontal scaling in the computation layer, and heterogeneous storage systems in the storage layers are three crucial challenges in a modern centralized SQL platform, focusing on interactive queries whose latencies range from seconds to minutes. We discussed the evolution of the hybrid-cloud SQL federation system in Twitter's data platform, aiming to address these challenges. The demonstrated hybrid-cloud SQL federation system overcomes these challenges by implementing query federation, cluster federation, and storage federation, with a unified logical view in each layer. 

We discussed some lessons we learned from developing, deploying, and operating the SQL system. We found some differences such as capacity planning and IAM in on-premises and cloud environments, which usually lead to extra engineering effort. Additionally, from the observation on tens of millions of interactive queries at Twitter, most queries can be completed in less than 1 minute, but around 10\% of queries can cost many more compute resources to complete. We saw the necessity of various solutions, such as load balancing and cluster scaling, to tackle the challenge. Finally, we observed that SQL is still one of the most widely used languages in data analytics. It has even been extended to the machine learning domain. We hope these key findings could provide some deeper insights for building a large-scale interactive query platform.

\section*{Acknowledgment}

Twitter's SQL federation system is a complicated project that has evolved for years. We would like to express our gratitude to everyone who has served on Twitter's Interactive Query team, including former team members Hao Luo, Yaliang Wang, Da Cheng, Fred Dai, and Maosong Fu. We also appreciate Prateek Mukhedkar, Vrushali Channapattan, Daniel Lipkin, Derek Lyon, Srikanth Thiagarajan, Jeremy Zogg, and Sudhir Srinivas for their strategic vision, direction, and support to the team. Finally, we thank Erica Hessel, Alex Angarita Rosales, and the anonymous ECSA reviewers for their informative comments, which considerably improved our paper.

%
%
%
\bibliographystyle{splncs04}
\bibliography{library}

\end{document}